\documentclass[preprint,showpacs,showkeys, preprintnumbers,amsmath,amssymb,superscriptaddress, prl]{revtex4-1}

\usepackage{graphicx}
\usepackage{dcolumn}
\usepackage{bm}
\usepackage{braket}

\begin{document}


\title{Inter-subband Landau level couplings induced by in-plane magnetic fields in trilayer graphene}

\author{Yuta Asakawa}
\affiliation{Institute of Industrial Science, University of Tokyo,  4-6-1 Komaba, Meguro-ku, Tokyo 153-8505, Japan.}

\author{Satoru Masubuchi}
\email{msatoru@iis.u-tokyo.ac.jp}
\affiliation{Institute of Industrial Science, University of Tokyo,  4-6-1 Komaba, Meguro-ku, Tokyo 153-8505, Japan.}

\author{Naoko Inoue}
\affiliation{Institute of Industrial Science, University of Tokyo,  4-6-1 Komaba, Meguro-ku, Tokyo 153-8505, Japan.}

\author{Sei Morikawa}
\affiliation{Institute of Industrial Science, University of Tokyo,  4-6-1 Komaba, Meguro-ku, Tokyo 153-8505, Japan.}

\author{Kenji Watanabe}
\affiliation{Advanced Materials Laboratory, National Institute for Materials Science, 1-1 Namiki, Tsukuba, 305-0044, Japan. }

\author{Takashi Taniguchi}
\affiliation{Advanced Materials Laboratory, National Institute for Materials Science, 1-1 Namiki, Tsukuba, 305-0044, Japan. }

\author{Tomoki Machida}
\email{tmachida@iis.u-tokyo.ac.jp}
\affiliation{Institute of Industrial Science, University of Tokyo,  4-6-1 Komaba, Meguro-ku, Tokyo 153-8505, Japan.}

\begin{abstract}
We observed broken-symmetry quantum Hall effects and level crossings between spin- and valley- resolved Landau levels (LLs) in Bernal stacked trilayer graphene. When the magnetic field was tilted with respect to sample normal from $0^{\circ}$  to $66^\circ$, the LL crossings formed at intersections of zeroth and second LLs from monolayer-graphene-like and bilayer-graphene-like subbands, respectively, exhibited a sequence of transitions. The results indicate the LLs from different subbands  are coupled by in-plane magnetic fields ($B_{\parallel}$), which was explained by developing the tight-binding model Hamiltonian of trilayer graphene under $B_{\parallel}$.
\end{abstract}

\maketitle
The electronic properties of trilayer graphene (TLG) provides distinct multiple energy spectrum, consisting of overlapping monolayer-like linear and bilayer-like parabolic subbands \cite{Craciun2009,Taychatanapat2011,Bao2011,Henriksen2012,Zou2013,Lee2016,Berciaud2014}. The subbands in TLG derives from multiple atomic sites in a unit cell \cite{Koshino2009A}, which have different origin from those of the conventional semiconductor quantum well (QW), where multiple subbands derive from electrostatic confinement of electrons in QW \cite{Ando1982}. The quantum number to index multiple subbands in TLG is atomic sites of graphene lattice $\left(A_x,B_x\right)$ where $x$ is an layer number \cite{Zhang2012}, and the corresponding number in QW is the wavenumber of plane wave \cite{Ando1982}. In the previous experimental studies in semiconductor QW, in-plane magnetic fields $(B_\parallel)$ have been shown to induce intermixing effect between subbands \cite{Yoon2000,Schlesinger1983}, and are described in the framework of Schr\"odinger equation of continuum Fermi sea \cite{Ando1982}. In case of TLG, the wavefunctions in monolayer-like and bilayer-like bands consists of hybridized electronic states in layer-asymmetric states (LASs) $\left[A_1-A_3,B_1-B_3\right]$ and layer symmetric states (LSSs) $\left[A_1+A_3,B_2,A_2,B_1+B_3 \right]$, and their energy spectrum is described by the tight-binding model \cite{Koshino2009A,Guinea2006,Partones2006,Koshino2010A,Avetisyan2010}. Therefore, whether application of $B_\parallel$  affects intermixing between monolayer-like and bilayer-like subbands cannot be treated by the extension of conventional semiconductor QW theory.

In order to investigate the intermixing effect between MLG-like and BLG-like bands, one can utilize the crossing between Landau levels (LLs) in MLG-like and BLG-like bands, which scale as $\sqrt{B_\perp}$  and $B_\perp$, respectively. The previous studies revealed that application of perpendicular electric fields $E_\perp$ induced symmetry breaking between mirror symmetric states $A_1-A_3 \leftrightarrow A_1+A_3$ and $B_1-B_3 \leftrightarrow B_1+B_3$  , and LL anticrossing between the $n$-th LL from the MLG-like band and $(n+3)$-th LL from the BLG-like band for the $K^-$ valley ($(n+4)$-th LL from the BLG-like band for the $K^+$ valley) \cite{Stepanov2016,Serbyn2013, Campos2016, Shimazaki2016, Koshino2010B, Lu2006, Avetisyan2009, Apalkov2012}. In contrast, the effects of $B_\parallel$ in TLG have yet been investigated. Theoretical treatment of $B_\parallel$ is limited to double-layer system, which is described by combination of double monolayer graphene \cite{Pershoguba2015, Pershoguba2010}. The experimental studies have been conducted only in single-band materials, such as organic conductors \cite{Singleton2000}, and intercalated graphite \cite{Enomoto2006,Iye1994}. In this work, we report on the magnetotransport measurements of trilayer graphene, and show that $B_\parallel$ induces LL couplings between $n=0$ and $n=2$ LLs from MLG-like and BLG-like bands, respectively. These effects are explained by introducing coupling terms in tight-binding Hamiltonian which connects different sets of wavefunctions in LSSs and LASs compared to $E_\perp$.

We exfoliated graphene and hBN flakes on  SiO$_2$/Si wafer \cite{Novoselov2005A,Novoselov2004,Novoselov2005B} and assembled them into hBN/TLG/hBN stacks using the dry pick-up method \cite{Wang2013}. The Hall-bar geometry and Au/Pd/Cr (45/15/10 nm) metal contacts were fabricated using electron-beam lithography [Fig. 1(a)]. Transport measurements were conducted in a dilution refrigerator with a base temperature of  $T = 100$ mK. The sample was tilted in magnetic fields; thus, perpendicular $(B_\perp)$ and in-plane $(B_\parallel)$ magnetic fields were tuned as $B_\perp = B_\textrm{tot} \cos \theta$  and $B_\parallel = B_\textrm{tot} \sin \theta$, respectively, where $\theta$ is the direction of magnetic field  $B_\textrm{tot}$  with respect to the sample normal. The longitudinal resistance $R_{xx}$ was measured with the alternating current $I_\textrm{ac} = 10$ nA. A silicon substrate was utilized as the global back gate to tune the charge-carrier density according to $n_e=C_\textrm{g} (V_\textrm{g}-V_0 )/e$, where $C_\textrm{g}=9 \times 10^{-9}$  F/cm$^2$ is the gate capacitance, $V_\textrm{g}$ is the back-gate bias voltage, and $V_0$ is the value of $V_\textrm{g}$ at charge-neutrality point.

Figure 1(b) shows $R_{xx}$ vs. $V_\textrm{g}$ measured at $T=2$ K. The narrow peak and high mobility of $\mu \sim 1,200,000$ cm$^2$ V$^{-1}$ s$^{-1}$ indicated the unprecedented quality of our device. Figure 1(c) shows $R_{xx}$ as a function of  $B_\perp$  and $\nu$. Here, the value of $\nu$ was obtained as $\nu=n_e h/eB_\perp$, where $h$ is Planck's constant. $R_{xx}$ minima, represented by the blue stripes in Fig. 1(a), were observed at all integer  $\nu$ in the range of $-14 \leq \nu \leq 14$, indicating the complete lifting of spin and valley degeneracies of LLs. Under an intermediate magnetic field of  $1.5$ T$<B_\perp<6.0$ T, the $R_{xx}$ minima were disappeared in the regions indicated by dashed squares in Fig. 1(c). To characterize the feature, we calculated the LL spectrum of TLG by using the Slonczewski-Weiss-McClure parametrization of the tight-binding model, which contains seven hopping parameters  $(\gamma_0, \cdots, \gamma_5, \delta)$ \cite{Koshino2009A}. When the parameters were tuned as $\gamma_0=3.23$ eV, $\gamma_1=0.39$ eV, $\gamma_2=-0.0237$ eV , $\gamma_3=0.315$ eV , $\gamma_4=0.0438$ eV, $\gamma_5=0.006$ eV, and $\delta=0.0143$ eV, and estimating average displacement field $\Delta_1$ generated by the back-gate electrode at the LL crossing by the empirical relation $\Delta_1=e(U_1-U_3 )/2 \sim Ed/6=5.8$ \cite{Koshino2010B, Stepanov2016}, the calculated LL crossings [dashed squares in Fig. 1(d)] well reproduced the disappearance points of  minima [red dashed squares in Fig. 1(c)], from which we can attribute the observed QHE to those of Bernal stacked TLG.

The salient features in Fig. 1(c) are that the region of suppressed $R_{xx}$ is divided into several ring-like structures at $2<\nu<9$, as shown in Fig. 2(a). From the calculated LL spectrum [Fig. 2(b)], these features are attributed to the crossings between the spin and valley resolved $N=0$ and $N=2$ LLs from MLG-like ($0^\textrm{ml}$) and BLG-like ($2^\textrm{bl}$) bands [Fig. 2(b)]. In Fig. 2(a), we discerned 15 LL crossings [red circles] out of the expected $4 \times 4=16$, and they were attributed to those indicated by red circles in Fig.2 (b). Here, one LL crossing was missing along $\nu=9$ [gray circle in Fig. 2(b)], which may originate from QH ferromagnetism \cite{Datta2016}. In this work, we focus our attention on the remaining LL crossings. Here, as shown in Fig. 2(b), the spacing between LLs were significantly smaller than those between orbital-resolved LLs. In such situation the $R_{xx}$ is susceptible to slight changes in the LL structure. We varied $\theta$ to address the effects of $B_\parallel$ on LL structures.

Figures 3(a) shows color plots of $R_{xx}$ as a function of $B_\perp$ and $\nu$ measured at $\theta = 0^\circ, 20^\circ, 40^\circ,$ and $66^\circ$ [left to right]. When $\theta$ was increased, the region of suppressed $R_{xx}$ at $\nu = 3-9$ exhibited transitions [Fig. 3(a)]. Here, the presence of QHS was defined by the appearance of local minima in $R_{xx}$ vs. $\nu$ curves at each $B_\perp$, and we plotted the positions of QHS in the $B_\perp - \nu$ plane [Figs. S1 and S2 in the supplementary information].  Note that the significant changes of LL crossing structures were caused by $\theta$. When $\theta$ was increased, the region of QHS formed along $\nu = 3, 4, 5, 6,$ and $7$ were gradually extended, and the number of LL crossings was decreased from 15 to 4. In order to capture this behavior in detail, we show line cuts of Figs. 3(a) at varying $B_\perp$ for $\nu = 5, 6, $ and $7$ [Figs. 3(b)]. The positions of QHS are indicated by purple, blue, and green stripes in Figs. 3(c). At $\theta = 0^\circ$, $R_{xx}$ minima along $\nu=5$ were divided into two sets of $B_\perp$ as 4.3 T$< B_\perp<$ 4.6 T and, 4.9 T $<B_\perp$ and three LL crossings were observed at $B_\perp = 4.3$ T, $4.6$ T, and $4.9$ T. When $\theta$ was increased to $\theta = 20^\circ$, the QHS at $B_\perp > 4.9$ T were extended to smaller magnetic fields as $B_\perp > 4.8$ T  [Fig. 3(b)], and for further increase in $\theta$, the QHS were connected at $\theta = 40 ^\circ$. Finally, at $\theta = 66^\circ$ the QHS were developed for the entire range of $B_\perp$. QHS at $\nu=6$ exhibited different behavior from those at $\nu=5$. At $\theta = 0^\circ$, four LL crossings were present at $B_\perp = 4.2$ T, $4.4$ T, $4.9$ T, and $5.4$ T. On increasing $\theta$ from $0^\circ$ to $20^\circ$, two LL crossings at $B_\perp=4.2$ T and $4.4$ T were merged [Fig. 3(c)]. On further increasing $\theta$ to $40^\circ$, the LL crossings at $B_\perp = 4.9$ T and $5.4$ T were merged at $B_\perp \sim 5.3$ T. However, two LL crossings were preserved up to $\theta = 66^\circ$. In the case of $\nu = 7$, the LL crossings exhibited similar behavior to those at $\nu = 5$. At $\theta = 0^\circ$, the QHS were divided into three regions as $B_\perp < 4.3$ T, $4.7$ T$<B_\perp<5.4$ T, and $5.5$ T $<B_\perp$, When $\theta$ was increased, the QHS formed at $B_\perp < 4.3$ T and $\theta = 0^\circ$ were extended to higher $B_\perp$, and finally at $\theta = 66^\circ$, the QHS were developed at $4.5$ T $< B_\perp$. These observations clearly indicate that, by increasing $\theta$, significant structural changes occurred in LL structures, and finite energy gaps were generated at LL crossings along $\nu=5$ and $7$. Note that these results constitute the first direct observation of LL anticrossing induced by $B_\parallel$.

In order to explain the observed LL anticrossing behavior, we extended the conventional tight-binding Hamiltonian of TLG \cite{Pershoguba2015, Pershoguba2010, Partoens2007,Orlita2009} and developed a Hamiltonian of TLG under $B_\parallel$. In the standard Slonczewski-Weiss-McClure parametrization, with the basis of $[A_1,B_1,A_2,B_2,A_3,B_3]$, the Hamiltonian of TLG can be described as 

\[ H = 
\begin{pmatrix} 
  U_1 & v_0 \pi^\dag & v_4\pi^\dag&v_3\pi&\gamma_2/2&0\\
  v_0\pi&\delta+U1&\gamma_1&-v_4\pi^\dag&0&\gamma_5/2\\
  v_4\pi&\gamma_1&\delta+U_2&v_0\pi^\dag&-v_4\pi&\gamma_1\\
  v_3\pi^\dag & -v_4\pi&v_0\pi& U_2& v_3\pi^\dag& -v_4\pi\\
  \gamma_2/2&0&-v_4\pi^\dag&v_3\pi&U_3&v_0\pi^\dag\\
  0&\gamma_5/2&\gamma_1&-v_4\pi^\dag&v_0\pi&\delta+U_3
\end{pmatrix}
+\zeta g \mu_\textrm{B} B_\textrm{tot} I 
\]
where $\pi = \hbar(\xi k_x + i k_y)$, $\hbar v_i = \frac{\sqrt{3}}{2}a \gamma_i$, $U_{m=1, 2, 3}$  is a static potential at each graphene layer, $\xi = \pm1$ is an index for $K_+$ and $K_-$ valleys, $\zeta=\pm 1$ is an index for up and down spins, $g\sim2$ is Lande's g-factor, $\mu_\textrm{B}$ is the Bohr magneton, and $I$ is a unit matrix. Here, we included the effects of magnetic fields by the vector potential ${\bf A}({\bf r}) = (0, B_\perp x - B_\parallel z, 0)$ and Peierls phase $\exp{ ( i \frac{e}{\hbar} \int^{\bf R_j}_{\bf R_i} {\bf A}({\bf R}) \cdot d {\bf r})}$ \cite{Footnote002}. When we took the basis as $\{ \frac{A_1-A_3}{\sqrt{2}},\frac{B_1-B_3}{\sqrt{2}},\frac{A_1+A_3}{\sqrt{2}},B_2,A_2,\frac{B_1+B_3}{\sqrt{2}}\}$, the Hamiltonian can be expressed as $H=H_0 (B_\perp )+H' (\Delta_1, B_\parallel)+\zeta g \mu_\textrm{B} B_\textrm{tot} I$, with $H_0(B_\perp) = \begin{pmatrix} H_\textrm{MLG} (B_\perp)& 0\\ 0& H_\textrm{BLG}(B_\perp)\end{pmatrix}$, where $H_\textrm{MLG}$ $(H_\textrm{BLG})$   represents monolayer-like (bilayer-like) LLs \cite{Koshino2009A}. Here, $H' (\Delta_1, B_\parallel)$ block is expressed as
\[
H' (\Delta_1, B_\parallel) = \begin{pmatrix} 0&H_{\Delta_1, \parallel} \\ H_{\Delta_1, \parallel}^\dag&0 \end{pmatrix}\textrm{\,\, and}
\]
\[
 H_{\Delta_1, \parallel} = \begin{pmatrix} \Delta_1&iB_\parallel d v_3 / \sqrt{2} & iB_\parallel d v_4 / \sqrt{2}& -iB_\parallel d v_0 \\ iB_\parallel d v_0 & i B_\parallel d v_4 / \sqrt{2}&0&\Delta_1 \end{pmatrix}.
\]

Note that, $H' (\Delta_1, B_\parallel)$ has off-diagonal terms which connect MLG-like and BLG-like energy bands ($H_{\Delta_1, \parallel}$), and $B_\parallel$ are introduced in a different positions compared to $\Delta_1$. 

By numerically diagonalizing the Hamiltonian presented above, we calculated LL spectrum with $H' (\Delta_1, B_\parallel)$ and $H' (\Delta_1, B_\parallel=0)$ [Figs. 4(a) and S3 in the supplementary information]. In case of $H' (\Delta_1, B_\parallel=0)$  only the Zeeman terms were increased with $\theta$; therefore, LLs were merely shifted, and no LL anticrossing was observed [Figs. S11(a)-(d) in the supplementary information]. Thus, the Hamiltonian without $B_\parallel$ terms fails to explain the observation in Fig. 3. When $H' (\Delta_1, B_\parallel)$ term was included, the LL spectrum exhibited significant changes [Fig. 4(a)]. When $\theta$ was increased to $\theta = 20^\circ$, small anticrossing gaps emerged at two sets of LL crossings $(0^\textrm{ml}_{K_+, \uparrow}, 2^\textrm{bl}_{K_+,\uparrow})$ and $(0^\textrm{ml}_{K_+, \downarrow}, 2^\textrm{bl}_{K_+,\downarrow})$, as indicated by the solid arrows in the second panel of Fig. 4(a). When $\theta$ was further increased to $\theta = 40^\circ$, anticrossing gaps were developed at the crossings between $(0^\textrm{ml}_{K_-, \uparrow}, 2^\textrm{bl}_{K_-,\uparrow})$  and  $(0^\textrm{ml}_{K_-, \downarrow}, 2^\textrm{bl}_{K_-,\downarrow})$ [third panel of Fig. 4(a)]. Finally, at $\theta = 66^\circ$ , the anticrossing behavior resulted in structural changes of the LL spectrum [fourth panel of Fig. 4(a)]. In Fig. 4(b), we show the separation between LLs $\Delta E_\nu$ at $\nu = 5-7$ with $H' (\Delta_1, B_\parallel=0)$ (dotted curves) and $H' (\Delta_1, B_\parallel)$ (solid curves). When  $\theta$ was increased, at $\nu = 5$ , the number of LL crossings decreased from 3 to 0. Similarly, at $\nu = 7$, the number of LL crossings decreased from 3 to 1. On the other hand, at $\nu = 6$ , 4 LL crossings that were formed at $\theta = 0^\circ$  decreased but two were retained up to $\theta = 66^\circ$. On comparing these results with the experimental data presented in Fig. 3, our model calculation well explains the observed features in QHS at $\nu = 5, 6$ and $7$. From these results, we conclude that the observed changes in QHE under $B_\parallel$ to LL anticrossing  induced by $H' (\Delta_1, B_\parallel)$. 

Here, we discuss the interaction between LLs induced by $H' (\Delta_1, B_\parallel)$ \cite{Footnote001} based on the approximated eigenwavefunctions of the tight-binding Hamiltonian. When neglecting the trigonal warping terms $v_3$, the eigenwavefunctions of MLG-like and BLG-like bands at $\{ K^+, K^-\}$ valleys can be written by the combination of harmonic oscillator states as $\ket{\textrm{ML},n,\pm} = \{(c_1\ket{n-1}, c_2 \ket{n})^T, (c'_1\ket{n},c'_2\ket{n-1})^T\}$ \\
and $\ket{\textrm{BL},n,\pm} = $ $\{(c_3\ket{n-2},c_4\ket{n},c_5\ket{n-1},c_6\ket{n-1})^T, $ \\ $ (c'_3\ket{n},c'_4\ket{n-2},c'_5\ket{n-1},c'_6\ket{n-1})^T\}$. Based on these eigenwavefunctions, the effects of $H' (\Delta_1, B_\parallel)$ can be stated as to couple the harmonic oscillators between LSSs and LASs. In case of $K^+$ valley, the inner product of $H' (\Delta_1, B_\parallel)$ by LLs $\ket{\textrm{ML},2,+}$ and $\ket{\textrm{BL},0,+}$ takes $iB_\parallel d v_0 c_3 c_2^*$. This implies that the electronic states between $c_2\ket{n}$ at $[A_1-A_3]$ and $c_3\ket{n-2}$ at $[A_1+A_3]$ are coupled through electron hopping by $\gamma_0$. In case of $K^-$ valley, the inner product of $H' (\Delta_1, B_\parallel)$ by LLs $\ket{\textrm{ML},2,-}$ and $\ket{\textrm{BL},0,-}$  becomes $B_\parallel d v_3 / \sqrt{2} c'_4 c_{1}^{'*}$, indicating the electronic states between $c'_1\ket{n}$ at $[A_1-A_3]$ and $c'_4\ket{n-2}$ at $[B_2]$ are coupled through electron hopping  by $\gamma_3$. The matrix elements for the $K^+$ valley $(v_0)$ was significantly larger than those for the $K^-$ valley $(v_3)$. These asymmetries are experimentally emerged as the difference in  the size of anticrossing gaps at $K^+$ valley [solid arrows in second panel of Fig. 4(a)] and $K^-$ valley [solid arrows in third panel of Fig. 4(a)]. Our Hamiltonian well explains observed LL anticrossing between $N=0$ LL from MLG-like bands and $N=2$ LL from BLG-like bands. Especially, the inner product of $H' (\Delta_1, B_\parallel)$ with $B_\parallel=0$ becomes zero, which indicate that $B_\parallel$ have qualitatively different effects from those originated from $E_\perp$.

Finally, we briefly comment on the LL crossing behaviors at other filling factors. At $\nu=3$ and $4$, the QHS were extended down to low magnetic fields with increasing $\theta$ [Figs. 3(a) and Fig. S2 in the supplementary information]. These results were well reproduced by our theoretical model calculation. At $\nu=9$, QHS were present throughout the range of $B_\perp$ studied at $\theta = 0^\circ$, and the theoretical model calculation predicts one LL crossing. These observations indicate that, even in the absence of in-plane magnetic fields, gaps with finite size were developed at the corresponding LL crossings. Considering that the LLs are spin- or valley-polarized in this region, the presence of a gap can indicate the emergence of ordered ground states in SU(4) QH systems around LL crossings in TLG, which was recently reported in \cite{Datta2016}. Our results, in combination with the states observed in Ref. \cite{Datta2016} can suggest that it can lead to the emergence of novel electronic ground states of the SU(4) QH system by tuning the interactions by $H' (\Delta_1, B_\parallel)$. 

In summary, we studied the magnetotransport properties of Bernal-stacked TLG under tilted magnetic fields. We observed anticrossing between the zeroth LL from monolayer-like band and second LL from bilayer-like bands. We developed a tight-binding Hamiltonian that accounts for the observed experimental results, indicating that the LL anticrossing behavior is induced by in-plane magnetic fields. Our observation indicates that application of $B_\parallel$ induced coupling between monolayer-like and bilayer-like bands in a different manner than previously studied $E_\perp$. This study opens a new tuning strategy for controlling the electronic ground states of TLG.

\begin{acknowledgments}
The authors acknowledge Mikito Koshino, Rai Moriya, Yusuke Hoshi, and Miho Arai for valuable discussions and technical assistance. This work was supported by the following: the Core Research for Evolutional Science and Technology (CREST), the Japan Science and Technology Agency (JST); JSPS KAKENHI Grant Numbers JP16H00982, JP25107003, JP25107004, and JP26248061; the JSPS Research Fellowship for Young Scientists.; the Project for Developing Innovation Systems of the Ministry of Education, Culture, Sports, Science, and Technology (MEXT).
\end{acknowledgments}

\clearpage
\begin{figure}[]
\begin{center}
\includegraphics[width=8.0 cm]{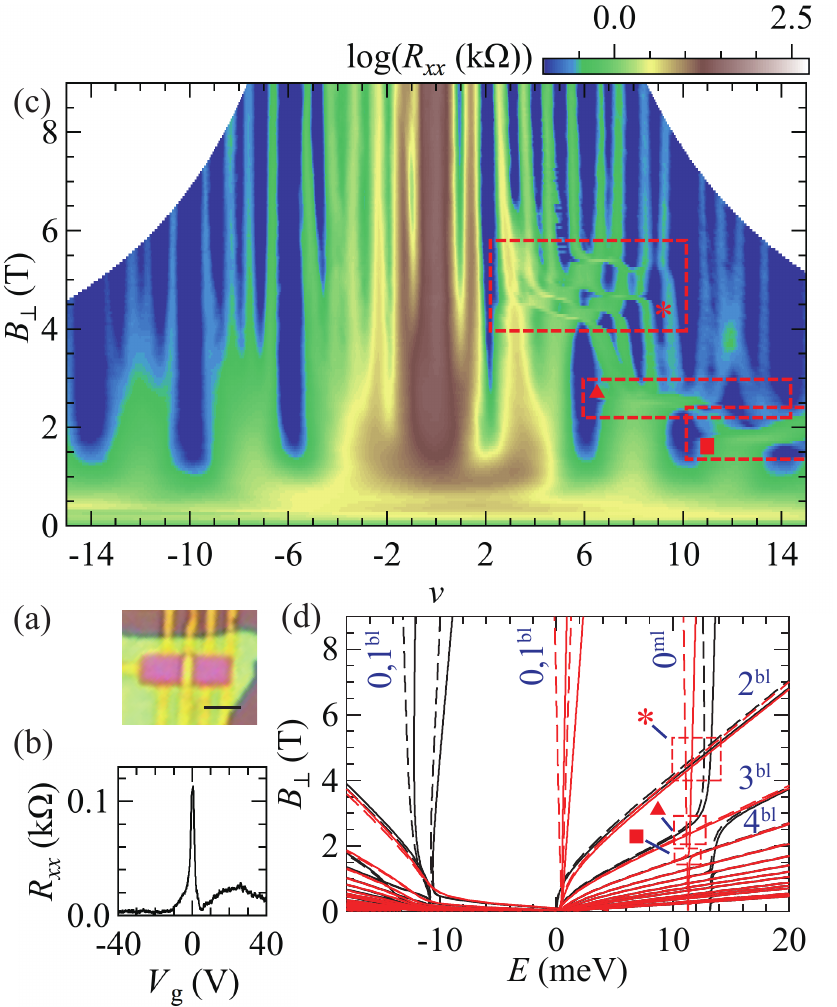} 
\caption { (color online) (a) Optical microscopy image of the device studied. The scale bar corresponds to 2 $\mu$m. (b) $R_{xx}$ vs. $V_\textrm{g}$ measured at 2 K. (c) Color plot of $R_{xx}$ as a function of $B_\perp$ and $\nu$ measured at 100 mK. (d) Calculated LL spectrum as a function of $B_\perp$. The red (black) curves correspond to $K^+$ $(K^-)$ valleys. Solid (dashed) curves indicate up (down) spins.}
\end{center}
\end{figure}

\clearpage
\begin{figure}[]
\begin{center}
\includegraphics[width=8.0 cm]{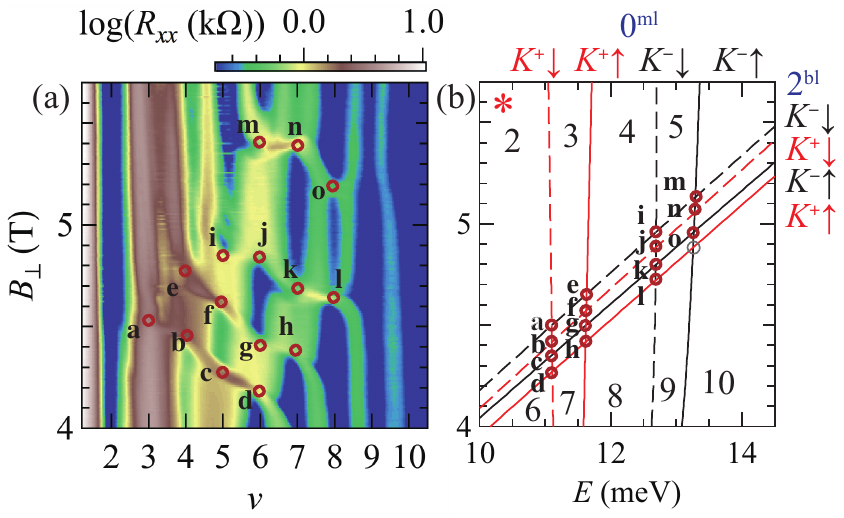} 
\caption {(color online) (a) Color plot of $R_{xx}$ as a function of $B_\perp$ and $\nu$. Red solid circles indicate the LL crossings. (b) Calculated LL spectrum at LL crossings. The red (black) curves correspond to $K^+$ $(K^-)$ valleys. Solid (dashed) curves indicate up (down) spins. The numbers indicate $\nu$ at corresponding LL gaps.}
\end{center}
\end{figure}

\clearpage
\begin{figure}[]
\begin{center}
\includegraphics[width=8.0 cm]{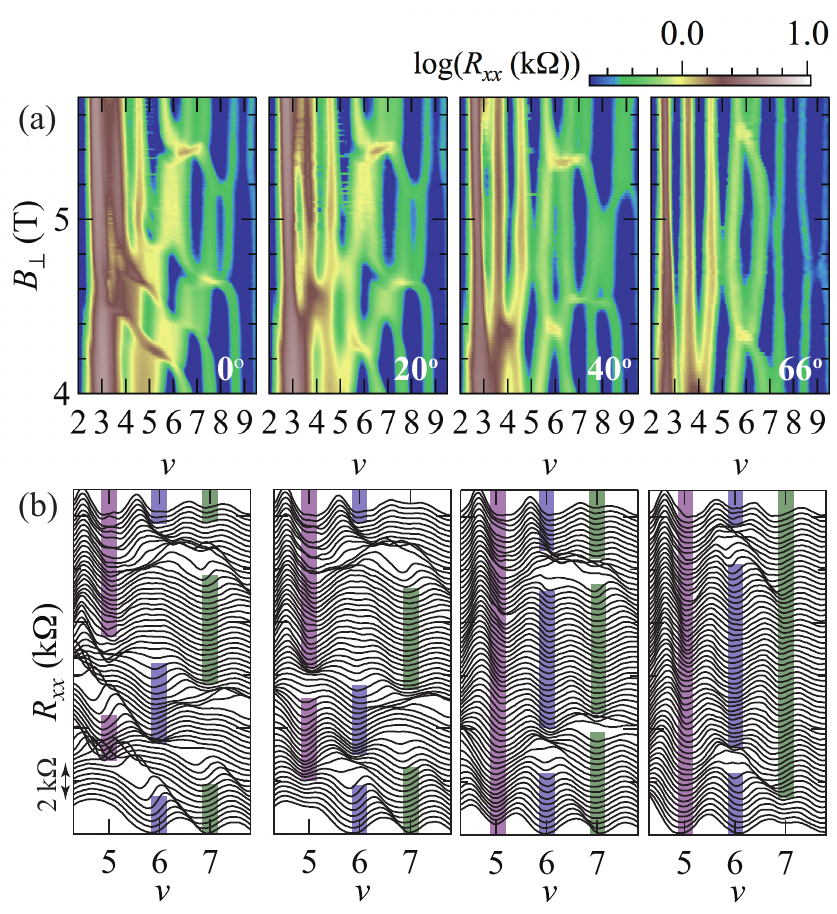} 
\caption { (color online) (a) Color plots of $R_{xx}$ as a function of $B_\perp$ and $\nu$ measured at varying $\theta=0^\circ, 20^\circ, 40^\circ,$ and $66^\circ$ (from left to right). (b) Line cuts of (a) between $\nu = 5$ and $7$. Each curve was offset vertically. The bottom (top) curves were measured at $B_\perp = 4.0$ T ($B_\perp = 5.7$ T). The color bars overlaid on these plots indicate the ranges of $B_\perp$ where QHS emerged.  }
\end{center}
\end{figure}

\clearpage
\begin{figure*}[]
\begin{center}
\includegraphics[width=16.0 cm]{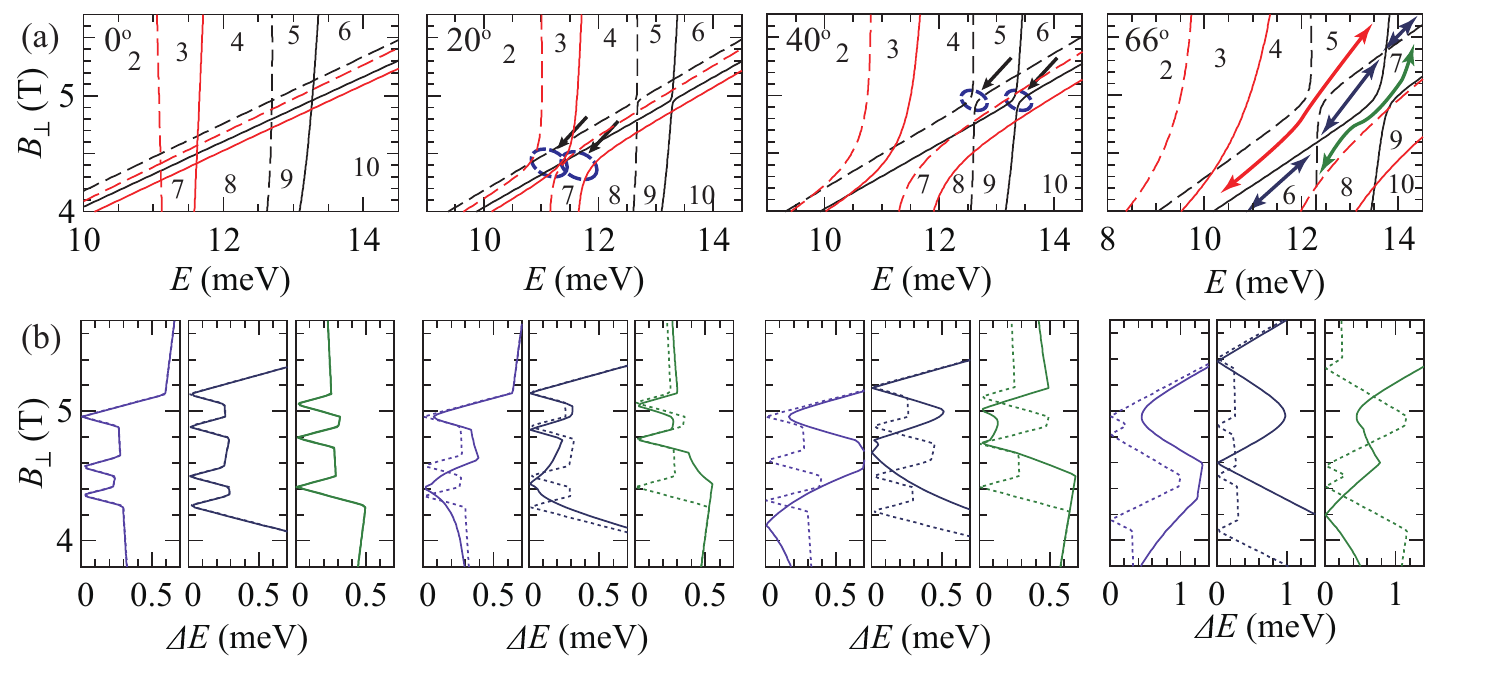} 
\caption {(color online) (a) LL spectrum as a function of $B_\perp$ for $\theta=0^\circ, 20^\circ, 40^\circ,$ and $66^\circ$ (from left to right) calculated using $H' (\Delta_1, B_\parallel)$. The red (black) curves indicate $K^+$ $(K^-)$  valleys. Solid (dashed) curves indicate up (down) spins. The figures indicate $\nu$ at corresponding LL gaps. At $\theta = 66^\circ$ LL gaps at $\nu=5, 6, $and $7$ are highlighted by red, blue, and green arrows, respectively. (b) The values of LL gaps $\Delta E$ as a function of $B_\perp$  at $\nu=$ (purple) $5$, (blue) 6, (green) 7 for   $\theta=0^\circ, 20^\circ, 40^\circ,$ and $66^\circ$  (from left to right), calculated by  $H' (\Delta_1, B_\parallel=0)$  (dotted curves) and $H' (\Delta_1, B_\parallel)$ (solid curves), respectively.  }
\end{center}
\end{figure*}

\end{document}